\documentclass{article}

\usepackage{arxiv}

\usepackage[utf8]{inputenc} % allow utf-8 input
\usepackage[T1]{fontenc}    % use 8-bit T1 fonts
\usepackage{hyperref}       % hyperlinks
\usepackage{url}            % simple URL typesetting
\usepackage{booktabs}       % professional-quality tables
\usepackage{amsfonts}       % blackboard math symbols
\usepackage{nicefrac}       % compact symbols for 1/2, etc.
\usepackage{microtype}      % microtypography
\usepackage{lipsum}		% Can be removed after putting your text content
\usepackage{graphicx}
\usepackage{adjustbox}
\usepackage{multirow}
\usepackage{siunitx}
\usepackage{caption}
\usepackage{threeparttable}
\usepackage{wrapfig}
\title{K-EmoCon, a multimodal sensor dataset for continuous emotion recognition in naturalistic conversations}

\date{} 					% Or removing it

\author{
    Cheul Young Park\textsuperscript{$*$}, Narae Cha, Soowon Kang, Auk Kim, Uichin Lee\thanks{Correspondence and requests for materials should be addressed to C.P. (email: \url{cheulyop@kaist.ac.kr}) or U.L. (email: \url{uclee@kaist.edu})} \\
	Graduate School of Knowledge Service Engineering, KAIST \\
	\texttt{\{cheulyop, nr.cha, sw.kang, kimauk, uclee\}@kaist.ac.kr} \\
	\And
	Ahsan Habib Khandoker, Leontios Hadjileontiadis \\
	Department of Biomedical Engineering, Khalifa University \\
	\texttt{\{ahsan.khandoker, leontios.hadjileontiadis\}@ku.ac.ae } \\
	\AND
	Alice Oh \\
	Department of Computer Science, KAIST \\
	\texttt{alice.oh@kaist.ac.kr} \\
	\And
	Yong Jeong \\
	Department of Bio and Brain Engineering, KAIST \\
	\texttt{yong@kaist.ac.kr} \\
	\And
}

\begin{document}
\maketitle

% ----------------------------------------------------------------
\begin{abstract}
Recognizing emotions during social interactions has many potential applications with the popularization of low-cost mobile sensors, but a challenge remains with the lack of naturalistic affective interaction data. Most existing emotion datasets do not support studying idiosyncratic emotions arising in the wild as they were collected in constrained environments. Therefore, studying emotions in the context of social interactions requires a novel dataset, and K-EmoCon is such a multimodal dataset with comprehensive annotations of continuous emotions during naturalistic conversations. The dataset contains multimodal measurements, including audiovisual recordings, EEG, and peripheral physiological signals, acquired with off-the-shelf devices from 16 sessions of approximately 10-minute long paired debates on a social issue. Distinct from previous datasets, it includes emotion annotations from all three available perspectives: self, debate partner, and external observers. Raters annotated emotional displays at intervals of every 5 seconds while viewing the debate footage, in terms of arousal-valence and 18 additional categorical emotions. The resulting K-EmoCon is the first publicly available emotion dataset accommodating the multiperspective assessment of emotions during social interactions.
\end{abstract}

% keywords can be removed
% \keywords{First keyword \and Second keyword \and More}

% ----------------------------------------------------------------
\section{Background \& Summary}

Emotion recognition research seeks to enable computers to identify emotions. It is a foundation for creating machines capable of understanding emotions, and possibly, even expressing one. Such a set of skills to recognize, understand, and express emotions form emotional intelligence~\cite{salovey1990emotional,mayer1999emotional}. It is suggested that emotional intelligence is necessary for the navigation of oneself within a society, as it allows one to reason what is desirable and what is not, and to regulate behaviors of self and others accordingly~\cite{salovey1997emotional,lopes2004emotional}.

Then why do machines need emotional skills? With advances in Machine Learning and Artificial Intelligence, the transition from human to machine is noticeable in all areas of the society, including those requiring expertise such as medical prognosis/diagnosis~\cite{esteva2017dermatologist,mastoras2019touchscreen} or automobile driving~\cite{cbinsight2020automotive}. It seems inevitable that these narrow AI systems~\cite{pennachin2007contemporary} supersede human experts in respective domains, as it has already been demonstrated with AlphaGo's superior performance in the game of Go over human champions~\cite{silver2016mastering,silver2017mastering}.

Not all AI will compete with humans, albeit their superhuman ability. Instead, many AI systems will work with us or for us. Emotional intelligence is critical for such human-computer interaction systems~\cite{reeves1996media}. Imagine a smart speaker that delightfully greets users when they come home. How should a speaker greet when a user had a rough day? A speaker neglectful of the user's emotional states may aggravate the user, but a speaker aware of the user's temper could remain silent to avoid the trouble. Similarly, emotional intelligence is critical for AI systems designed for complex tasks. For example, on roads where autonomous and human-driven vehicles mix, accurate recognition of emotions of human drivers' by autonomous vehicles would lead to more safety as autonomous vehicles can better judge human drivers' intentions~\cite{turpen2019selfdriving}.

Now for machines to become emotionally intelligent, they must first learn to recognize emotions, and the prerequisite to learning is data. However, there lie several challenges in the acquisition of emotion data. While emotions are prevalent, their accurate measurement is difficult. Most commonly, emotions are viewed as psychological states expressed through faces, with distinct categories~\cite{barrett2017emotions}, but research evidence claims the contrary. Rather than distinct, facial expressions are compound~\cite{du2014compound}, relative~\cite{yannakakis2017ordinal}, and misleading~\cite{frank2015microexpressions}. A recent review of scientific evidence also presses against the common view, suggesting that facial expressions lack reliability, specificity, and generalizability~\cite{barrett2019emotional}, together with past studies on contextual dependency~\cite{carroll1996facial,cauldwell2000did,barrett2011context} and individual variability of emotions~\cite{larsen1987affect,gross2003individual}.

Such inherent elusiveness of emotion renders many existing emotion datasets inapplicable for studying emotions in the wild. The majority of emotion datasets consist of emotions induced with selected stimuli in a static environment, i.e., a laboratory~\cite{soleymani2011multimodal,koelstra2011deap,abadi2015decaf,subramanian2016ascertain,katsigiannis2017dreamer,correa2018amigos,sharma2019dataset}. This method provides experimenters with full-control over data collection, allowing assessment of specific emotional behaviors~\cite{yan2013casme,schmidt2018introducing} and acquiring fine-grained data with advanced techniques like neuroimaging. Nevertheless, lab-generated data may generalize poorly to realistic scenarios as they frequently contain intense expressions of prototypical emotions, which are rarely observed in the real world~\cite{watson2000mood,batliner2003find}, acquired from only a subset of the population~\cite{henrich2010weirdest}.

An alternative approach utilizes media contents~\cite{dhall2012collecting,mollahosseini2017affectnet,mcduff2018fed+,poria2019meld} and crowdsourcing~\cite{mcduff2012crowdsourcing}, compensating for the shortcomings of the conventional method. The abundance of contents available online, such as TV-shows and movies, allows researchers to glean rich emotion data representative of various contexts efficiently. Crowdsourcing further supports inexpensive data annotation while serving as another data source~\cite{morris2014crowdsourcing,korovina2019reliability}. Datasets of this type have advantages in sample size and the diversity of subjects, but generalizability remains an issue. Datasets based on media contents often contain emotional displays produced by trained actors supposing fictitious situations. To what extent such emotional portrayals resemble spontaneous emotional expressions is debatable~\cite{motley1988facial,jurgens2015effect,juslin2018mirror}. They also provide no access to physiological signals, which are known to carry information vital for the detection of less visible changes in emotional states~\cite{cacioppo2000psychophysiology,picard2001toward,lisetti2004using,rainville2006basic,nummenmaa2014bodily,pace2019physiological}.

To amend this lack of a dataset for recognition of emotions in their natural forms, we introduce K-EmoCon, a multimodal dataset acquired from 32 subjects participating in 16 paired debates on a social issue. It consists of physiological sensor data collected with three off-the-shelf wearable devices, audiovisual footage of participants during the debate, and continuous emotion annotations. It contributes to the current literature of emotion recognition, as according to our knowledge, it is the first dataset with emotion annotations from all possible perspectives as the following: subject him/herself, debate partner, and external observers.

% ----------------------------------------------------------------
\section{Methods}
\subsection{Dataset design}

\paragraph{\textit{Intended usage}} Inspired by previous works that set out to investigate emotions during conversations~\cite{busso2008iemocap,mckeown2011semaine,busso2016msp,poria2019meld}, K-EmoCon was designed in consideration of a social interaction scenario involving two people and wearable devices capable of unobtrusive tracking of physiological signals. The dataset aims to allow a multi-perspective analysis of emotions with the following objectives:

\begin{enumerate}
    \item Extend the research on how having multiple perspectives on emotional expressions may improve their automatic recognition.
    \item Provide a novel opportunity to investigate how emotions can be perceived differently from multiple perspectives, especially in the context of social interaction.
\end{enumerate}

Previous research has shown that having multiple sources for emotion annotations can increase their recognition accuracy~\cite{healey2011recording,zhang2016automatic}. However, no research in our awareness employs all three available perspectives in the annotation of emotions (i.e., subject him/herself, interacting partner, and external observers). Having multiple perspectives relates to the issue of establishing ground truth in emotion annotations. Emotions are inherently internal phenomena, and their mechanism is unavailable for external scrutiny, even for oneself who is experiencing emotions. As a result, there may not be a ground truth for emotions. Should we consider what is most agreed upon by external observers of emotions as the ground truth, or what the person who experiences emotions reports to have felt the ground truth~\cite{truong2007unobtrusive}? Two views are likely to match if emotions are intense and pure, but as discussed, such emotions are rare. Instead, self-reported and observed emotions are likely to disagree for a variety of reasons. People often conceal their true emotions; sometimes, people are not fully mindful of their internal states; and some people may have difficulties in interpreting or articulating emotions~\cite{grossman2000verbal,dickson2014misperceptions}.

With K-EmoCon, we intend to enable the comprehensive examination of such cases where perceptions of emotions do not match, by bringing all three available perspectives into the annotation of emotions, in the context of a social interaction involving three parties of:

\begin{enumerate}
    \item \textit{The subject} -- is the source who experiences emotions firsthand and produces \textit{self annotations}, particularly the \textit{``felt sense''}~\cite{zhang2016automatic} of the emotions.
    \item \textit{The partner} -- is the person who interacts with the subject, experiencing the subject's emotions secondhand; thus, he or she has a contextual knowledge of the interaction that induced the subject's emotions and produces \textit{partner annotations} based on that.
    \item \textit{The external observers} -- are people who observe the subject's emotions without the exact contextual knowledge of the interaction that induced the emotions, producing \textit{external observer annotations}.
\end{enumerate}

Notice, that while our definition of perspectives involved in emotion annotation is similar to definitions previously used by other researchers (self-reported vs. perceived~\cite{zhang2016automatic}/observed~\cite{truong2012speech}), we further segment observer annotations based on whether the contextual information of the situation in which the emotion was generated is available to an observer, as we wish to consider the role of contextual knowledge in emotion perception and recognition.

Existing datasets of emotions in conversations provide a limited scope on this issue as they at most contain emotion annotations from subjects and external observers~\cite{busso2008iemocap}, leaving out annotations from other people who engaged in the conversation (whom we call partners). Or, they either only consider a particular type of annotations that is sufficient to serve their research goal~\cite{busso2016msp} or their designs do not allow acquiring multi-perspective annotations~\cite{mckeown2011semaine,poria2019meld} (e.g., a dataset is constructed upon conversations from a TV-show, only allowing the collection of external observer annotations). Refer to Table~\ref{table1:datasets} to see how K-EmoCon is distinguished from existing emotion datasets.

\begin{table}[!h]
\centering
\addtolength{\leftskip} {-2cm} % increase (absolute) value if needed
\addtolength{\rightskip}{-2cm}
\begin{threeparttable}
\caption{Comparison of the K-EmoCon dataset with the existing multimodal emotion recognition datasets. Posed emotions are when a subject is instructed to enact a particular emotion while Spon. = spontaneous. Similarly, induced emotions are when a set of selected stimuli is used for their elicitation. For annotation types, S = \textit{self annotations}, P = \textit{partner annotations}, and E = \textit{external observer annotations}.}
\renewcommand{\arraystretch}{1.2}
\begin{tabular}{@{}lclcclll@{}}
\toprule
\textbf{Name (year)} & \textbf{Size} & \textbf{Modalities} & \textbf{\begin{tabular}[c]{@{}l@{}}Spon. vs. \\ posed\end{tabular}} & \textbf{\begin{tabular}[c]{@{}l@{}}Natural vs. \\ induced\end{tabular}} & \textbf{\begin{tabular}[c]{@{}l@{}}Annotation \\ method\end{tabular}} & \textbf{\begin{tabular}[c]{@{}l@{}}Annotation \\ type\end{tabular}} & \textbf{Context} \\ \midrule
\begin{tabular}[c]{@{}l@{}}IEMOCAP\\ (2008)~\cite{busso2008iemocap}\end{tabular} & 10 & \begin{tabular}[c]{@{}l@{}}Videos, face motion \\ capture, gesture, speech \\ (audio \& transcribed)\end{tabular} & Both & Both\tnote{\dag} & \begin{tabular}[c]{@{}l@{}}Per dialog \\ turn\end{tabular} & S, E & Dyadic \\ \midrule
\begin{tabular}[c]{@{}l@{}}SEMAINE\\ (2011)~\cite{mckeown2011semaine}\end{tabular} & 150 & \begin{tabular}[c]{@{}l@{}}Videos, FAUs, speech \\ (audio \& transcribed)\end{tabular} & Spon. & Induced & \begin{tabular}[c]{@{}l@{}}Trace-style \\ continuous\end{tabular} & E & Dyadic \\ \midrule
\begin{tabular}[c]{@{}l@{}}MAHNOB-HCI\\ (2011)~\cite{soleymani2011multimodal}\end{tabular} & 27 & \begin{tabular}[c]{@{}l@{}}Videos (face and body), \\ eye gaze, audio, biosignals \\ (EEG, GSR, ECG, respiration, \\ skin temp.)\end{tabular} & Spon. & Induced & Per stimuli & S & Individual \\ \midrule
\begin{tabular}[c]{@{}l@{}}DEAP\\ (2012)~\cite{koelstra2011deap}\end{tabular} & 32 & \begin{tabular}[c]{@{}l@{}}Face videos, biosignals \\ (EEG, GSR, BVP, respiration, \\ skin temp., EMG \& EOG)\end{tabular} & Spon. & Induced & Per stimuli & S & Individual \\ \midrule
\begin{tabular}[c]{@{}l@{}}DECAF\\ (2015)~\cite{abadi2015decaf}\end{tabular} & 30 & \begin{tabular}[c]{@{}l@{}}NIR face videos, biosignals \\ (MEG, hEOG, ECG, tEMG)\end{tabular} & Spon. & Induced & Per stimuli & S & Individual \\ \midrule
\begin{tabular}[c]{@{}l@{}}ASCERTAIN\\ (2016)~\cite{subramanian2016ascertain}\end{tabular} & 58 & \begin{tabular}[c]{@{}l@{}}Facial motion units (EMO), \\ biosignals (ECG, GSR, EEG)\end{tabular} & Spon. & Induced & Per stimuli & S & Individual \\ \midrule
\begin{tabular}[c]{@{}l@{}}MSP-IMPROV\\ (2016)~\cite{busso2016msp}\end{tabular} & 12 & Face videos, speech audio & Both & Both\tnote{\dag} & \begin{tabular}[c]{@{}l@{}}Per dialog \\ turn\end{tabular} & E & Dyadic \\ \midrule
\begin{tabular}[c]{@{}l@{}}DREAMER\\ (2017)~\cite{katsigiannis2017dreamer}\end{tabular} & 23 & Biosignals (EEG, ECG) & Spon. & Induced & Per stimuli & S & Individual \\ \midrule
\begin{tabular}[c]{@{}l@{}}AMIGOS\\ (2018)~\cite{correa2018amigos}\end{tabular} & 40 & \begin{tabular}[c]{@{}l@{}}Vidoes (face \& body), \\ biosignals (EEG, ECG, GSR)\end{tabular} & Spon. & Induced & Per stimuli & S, E & \begin{tabular}[c]{@{}l@{}}Individual, \\ Group\end{tabular} \\ \midrule
\begin{tabular}[c]{@{}l@{}}MELD\\ (2019)~\cite{poria2019meld}\end{tabular} & 7 & \begin{tabular}[c]{@{}l@{}}Videos, speech \\ (audio \& transcribed)\end{tabular} & Both & Both\tnote{\dag} & Turn-based & E & \begin{tabular}[c]{@{}l@{}}Dyadic, \\ Group\end{tabular} \\ \midrule
\begin{tabular}[c]{@{}l@{}}CASE\\ (2019)~\cite{sharma2019dataset}\end{tabular} & 30 & \begin{tabular}[c]{@{}l@{}}Biosignals (ECG, respiration, \\ BVP, GSR, skin temp., EMG)\end{tabular} & Spon. & Induced & \begin{tabular}[c]{@{}l@{}}Trace-style \\ continuous\end{tabular} & S & Individual \\ \midrule
\begin{tabular}[c]{@{}l@{}}CLAS\\ (2020)~\cite{markova2019clas}\end{tabular} & 64 & \begin{tabular}[c]{@{}l@{}}Biosignals (ECG, PPG, EDA),\\ accelerometer\end{tabular} & Spon. & Induced & Per stimuli/task & Predefined\tnote{\ddag} & Individual \\ \midrule
\textit{\begin{tabular}[c]{@{}l@{}}K-EmoCon\\ (2020)\end{tabular}} & \textit{32} & \textit{\begin{tabular}[c]{@{}l@{}}Videos (face, gesture), \\ speech audio, accelerometer, \\ biosignals (EEG, ECG, BVP, \\ EDA, skin temp.)\end{tabular}} & \textit{Spon.} & \textit{Natural} & \textit{\begin{tabular}[c]{@{}l@{}}Interval-based \\ continuous\end{tabular}} & \textit{S, P, E} & \textit{Dyadic} \\ \bottomrule
\end{tabular}
\begin{tablenotes}[online]\medskip
    \item[\dag] A dataset was considered to contain induced emotions if scripted interaction was involved in the data collection, even though no artificial stimuli (such as an emotion inducing video clip) was used.
    \item[\ddag] Predefined emotion categories of stimuli and success rates of participants in a set of purposefully selected cognitive tasks were used as ground-truth labels.
\end{tablenotes}
\label{table1:datasets}
\end{threeparttable}
\end{table}

\paragraph{\textit{Context of data collection}} In this regard, we chose a semi-structured, turn-taking debate on a social issue with randomly assigned partners as the setting for data collection. This setting is appropriate for collecting emotions that may naturally arise in a day, as it is similar to a social interaction that one could engage in a workplace.

Also, the setting is particularly suitable for studying the misperception of emotions. It is sufficiently formal and spontaneous as it involves randomly assigned partners. We expect such formality and spontaneity of the setting compelled participants to regulate their emotions in a socially appropriate manner, allowing us to observe less pronounced emotions from participants, which were more likely to be misperceived by their partners~\cite{hess1997intensity}.

\paragraph{\textit{Data collection apparatus}} Our choice of mobile, wearable, and low-cost devices to collect affective physiological signals together with audiovisual recordings, while primarily aims to make findings based on our data more reproducible and expandable, was also in consideration of our goal of investigating mismatches in perceptions of emotions in the wild. Research has shown that fusing implicit and explicit affective information can result in more accurate recognition of subtle emotional expressions from professional actors~\cite{ranganathan2016multimodal}. However, no work we are aware of has shown that a similar result can be achieved for subtle emotions collected from in-the-wild social interactions of individuals without professional training in acting. Therefore, our dataset provides an opportunity to examine if emotions of lower intensity, produced from non-actors during communication, can be recognized accurately.

It is also interesting to examine whether subtle emotions could signal instances where emotions are misperceived during communication if their accurate detection is possible. In the same vein, to what extent the intensity of emotions influences their decoding accuracy during a social interaction, where a broader array of contextual information is present, is also worth exploring. K-EmoCon could enable an in-depth investigation of such issues.

Further, we considered the use case of mobile and wearable technologies for facilitating emotional communication. Researchers are actively exploring the potential for using expressive biosignals collected via wearables to communicate one's emotional and psychological states with others~\cite{min2014biosignal,hassib2017heartchat,liu2017supporting,liu2019animo,liu2019effect}. Our dataset can contribute to the research of biosignal-based assistive technologies to enable affective communication by providing insights on when are apposite moments for communicating emotions.

% ----------------------------------------------------------------
\subsection{Ethics statement}

The construction of the K-EmoCon dataset was approved by the Korea Advanced Institute of Science and Technology (KAIST) Institutional Review Board. KAIST IRB also reviewed and approved the consent form, which contained information on the following: the purpose of data collection, data collection procedure, types of data to be collected from participants, compensation to be provided for participation, and the protocol for the protection of privacy-sensitive data.

Participants were given the same consent forms upon arriving at the data collection site and were asked to provide written consent after fully reading the form indicating that they are willing to participate in data collection. Since K-EmoCon is to be open to public access, a separate consent was obtained for the disclosure of the data that contains personally identifiable information (PII), which is the audiovisual footage of participants during debates, including their faces and voices. Participants were also notified that their participation is voluntary, and they can terminate the data collection at any. The resulting K-EmoCon dataset includes the audiovisual recordings of 21 participants, out of 32, who agreed to disclose their personal information, excluding the 11 who did not agree.

% ----------------------------------------------------------------
\subsection{Participant recruitment and preparation}

\begin{wraptable}{r}{5.5cm}
\centering
\caption{Participant pairs for debates.}
\begin{tabular}{@{}cccc@{}}
\toprule
\multicolumn{2}{l}{Participants} & \multicolumn{2}{l}{Gender and ages} \\ \midrule
P1 & P2 & M (25) & M (23) \\
P3 & P4 & M (36) & M (25) \\
P5 & P6 & M (22) & M (23) \\
P7 & P8 & M (22) & F (22) \\
P9 & P10 & M (21) & M (22) \\
P11 & P12 & M (22) & M (25) \\
P13 & P14 & M (22) & F (21) \\
P15 & P16 & M (30) & F (26) \\
P17 & P18 & M (21) & M (20) \\
P19 & P20 & M (21) & F (23) \\
P21 & P22 & M (25) & F (25) \\
P23 & P24 & M (22) & F (29) \\
P25 & P26 & F (26) & M (25) \\
P27 & P28 & F (24) & F (23) \\
P29 & P30 & F (23) & F (24) \\
P31 & P32 & M (24) & F (19) \\ \bottomrule
\end{tabular}
\label{table2:participants}
\end{wraptable}

32 participants were recruited between January and March of 2019. An announcement calling for participation in an experiment on ``emotion-sensing during a debate'' was posted on an online bulletin board of a KAIST student community. The post stated that participants would have a debate on the issue of accepting Yemeni refugees on Jeju Island of South Korea for 10 minutes. It also stated that the debate must be in English, and participants should be capable of speaking competently in English, but not necessarily at the level of a native speaker. Specifically, participants were required to have at least three years of experience living in an English-speaking country, or have achieved a score above criteria in any one of standardized English speaking tests as listed here: TOEIC speaking level 7, TOEFL speaking score 27, or IELTS speaking level 7.

Once participants were assigned a date and time to participate in data collection, they were provided four news articles on the topic of the Jeju Yemeni refugee crisis via email. The articles included two articles with neutral views on the issue~\cite{kim2018fundamental,kang2018major}, one in favor of refugees~\cite{park2018crazy}, and one in opposition to refugees~\cite{seo2018opposition}. We instructed the participants to read the articles beforehand to familiarize themselves with the debate topic.

All selected participants were students at KAIST, but their ages varied from 19 to 36 years old (mean = 23.8 years, stdev. = 3.3 years), as well as their gender and nationality. We randomly paired participants into 16 dyads based on their available times. See Table~\ref{table2:participants} for the breakdown of participants' gender, nationality, and age.

% ----------------------------------------------------------------
\begin{figure}[!t]
\centering
\includegraphics[width=\textwidth]{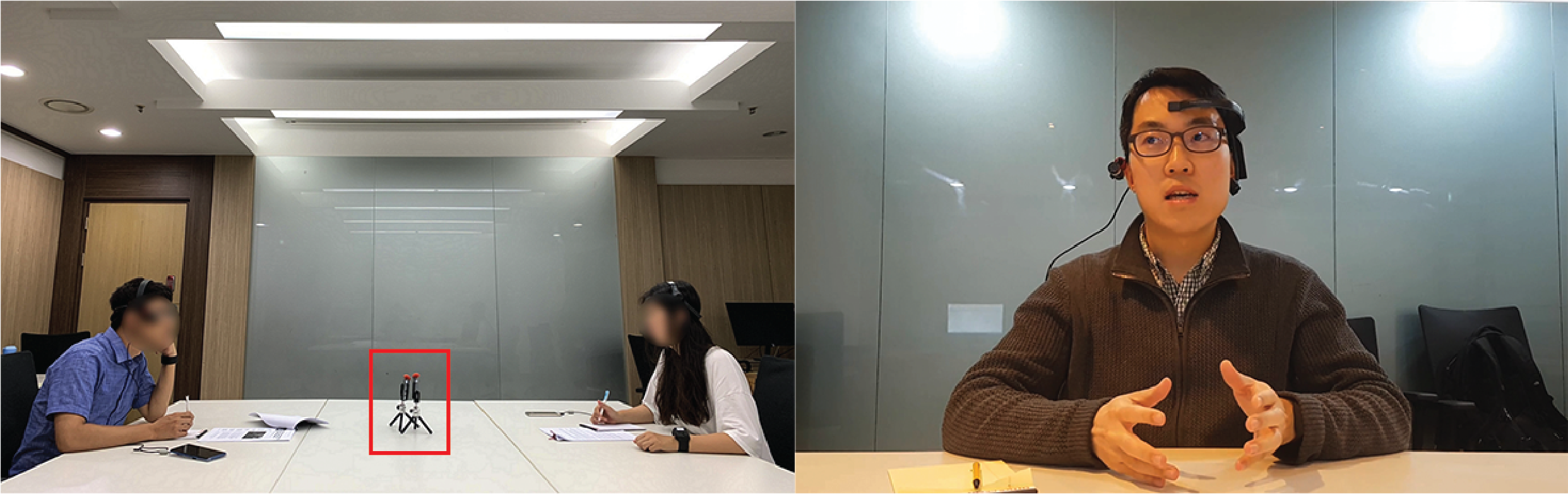}
\caption{Picture on the left shows a pair of participants sitting at a table preparing for a debate. Two smartphones on tripods in the middle of the table (highlighted in red) recorded participants' facial expressions and movements in their upper body, as shown on the right in the sample screenshot of footage.}
\label{fig1:setup}
\end{figure}

\subsection{Data collection setup}

All data collection sessions were conducted in two rooms with controlled temperature and illumination. Two participants sat across a table facing each other with a distance in between for a comfortable communication (see Figure~\ref{fig1:setup}). Two Samsung Galaxy S7 smartphones mounted on tripods were placed in the middle of the table facing each participant, capturing facial expressions and movements in the upper body from the 3rd-person point of view (POV) along with the speech audio, via the camera app.

\begin{figure}[!t]
\centering
\includegraphics[width=\textwidth]{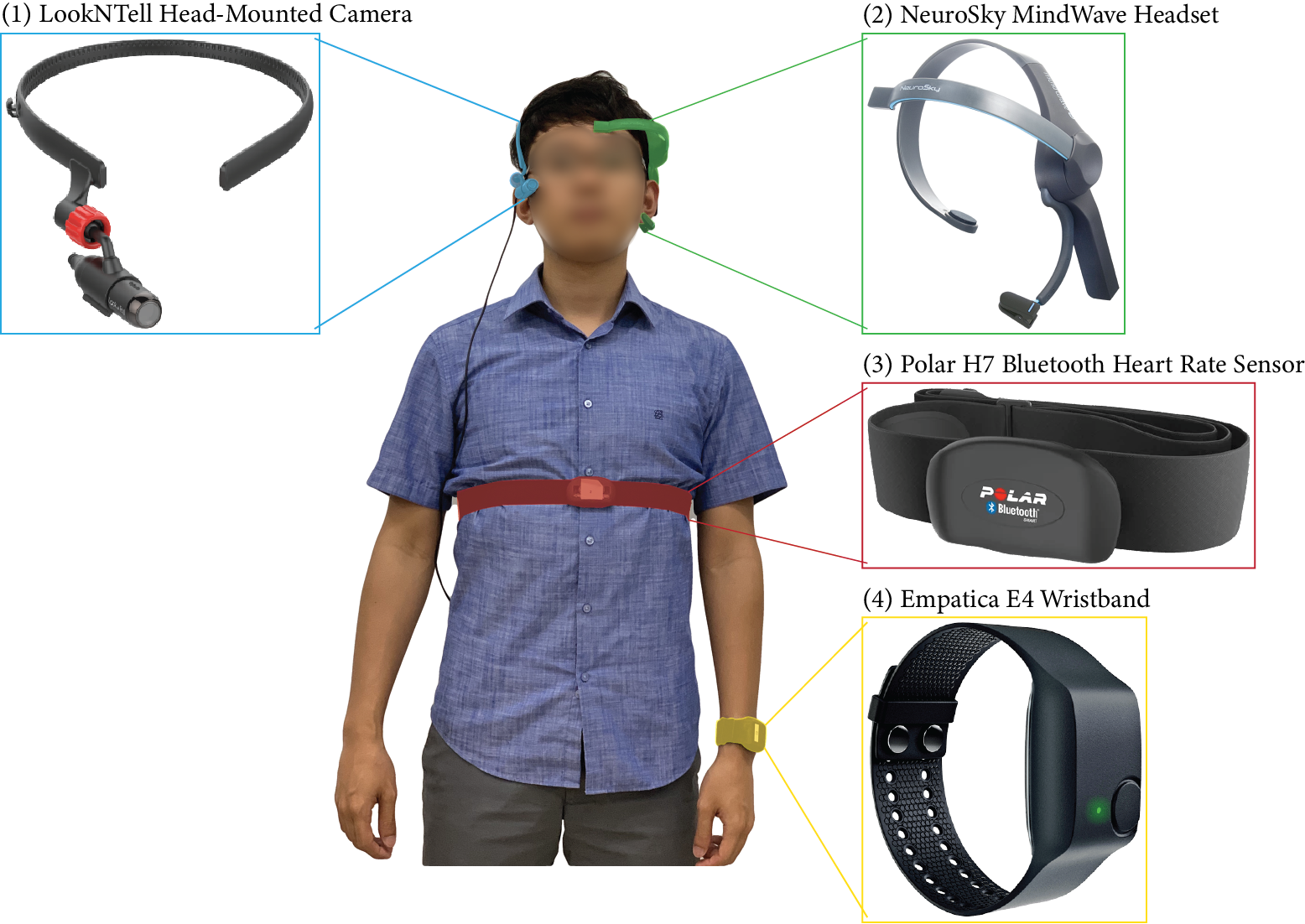}
\caption{Frontal view of a participant equipped with wearable sensors.}
\label{fig2:devices}
\end{figure}

During a debate, participants wore a suite of wearable sensors, as shown in Figure~\ref{fig2:devices}, which includes:

\begin{enumerate}
    \item \textit{Empatica E4 Wristband} -- captured photoplethysmography (PPG), 3-axis acceleration, body temperature, and electrodermal activity (EDA). Heart rate and the inter-beat interval (IBI) were derived from Blood Volume Pulse (BVP) measured by a PPG sensor.
    \item \textit{Polar H7 Bluetooth Heart Rate Sensor} -- detected heart rates using an electrocardiogram (ECG) sensor and was used to complement a PPG sensor in E4, which is susceptible to motion.
    \item \textit{NeuroSky MindWave Headset} -- collected electroencephalogram (EEG) signals via two dry sensor electrodes, one on the forehead (fp1 channel-10/20 system at the frontal lobe) and one on the left earlobe (reference).
    \item \textit{LookNTell Head-Mounted Camera} -- with a camera attached at one end of a plastic circlet, was worn on participants' heads to capture videos from a first-person POV.
\end{enumerate}

All listed devices can operate in a mobile setting. Empatica E4 keeps the data on the device, and the collected data is later uploaded to a computer. Polar H7 sensor and MindWave headset can communicate with a mobile phone via Bluetooth Low Energy (BLE) to store data. Table~\ref{table3:devices} summarizes sampling rates and signal ranges of data collected from each device.

\begin{table}[!t]
\centering
\caption{Data collected with each wearable device, with respective sampling rates and signal ranges.}
\renewcommand{\arraystretch}{1.5}
\begin{tabular}{@{}llll@{}}
\toprule
\textbf{Devices} & \textbf{Collected data} & \textbf{Sampling rate} & \textbf{Signal range {[}min, max{]}} \\ \midrule
\multirow{6}{*}{Empatica E4 Wristband} & 3-axis acceleration & 32Hz & {[}-2g, 2g{]} \\ \cline{2-4}
 & BVP (PPG) & 64Hz & n/a \\ \cline{2-4}
 & EDA & 4Hz & {[}0.01$\mu$S, 100$\mu$S{]} \\ \cline{2-4}
 & Heart rate (from BVP) & 1Hz & n/a \\ \cline{2-4}
 & IBI (from BVP) & n/a & n/a \\ \cline{2-4}
 & Body temperature & 4Hz & {[}\SI{-40}{\celsius}, \SI{115}{\celsius}{]} \\ \hline
\multirow{2}{*}{NeuroSky MindWave Headset} & Brainwave (fp1 channel EEG) & 125Hz & n/a \\ \cline{2-4}
 & Attention \& Meditation & 1Hz & {[}0, 100{]} \\ \hline
Polar H7 Heart Rate Sensor & HR (ECG) & 2Hz & n/a \\ \bottomrule
\end{tabular}
\label{table3:devices}
\end{table}

% ----------------------------------------------------------------
\subsection{Data collection procedure}

\paragraph{\textit{Administration}} All data collection sessions were conducted in four stages of 1) onboarding, 2) baseline measurement, 3) debate, and 4) emotion annotation. Two experimenters administered each session (see Table~\ref{table4:protocol} for the overview of a data collection procedure). One experimenter served as a moderator during debates, notifying participants of the remaining time and intervening under any necessary circumstances, such as when a debate gets too heated, or a participant exceeds an allotted time of 2 minutes in his or her turn.

\paragraph{\textit{Onboarding}} Upon their arrival, participants were each provided a consent form asking for two written consents, first for the participation in data collection that was mandatory, and second for the disclosure of privacy-sensitive data collected during the session, which participants could opt-out without any disadvantage.

\begin{table}[!t]
\centering
\begin{threeparttable}
\caption{Steps for a data collection session.}
\renewcommand{\arraystretch}{1.2}
\begin{tabular}{@{}lll@{}}
\toprule
\textbf{Step} & \textbf{Allocated time} & \textbf{Description} \\ \midrule
Read and sign consent forms & 10 min & \begin{tabular}[c]{@{}l@{}}Experimenters provided consent forms to participants, and \\ two written consents each for participation and the collection \\ of privacy-sensitive data were obtained.\end{tabular} \\ \hline
Choose sides and the order & 5 min & \begin{tabular}[c]{@{}l@{}}Participants were assigned to either argue in favor of or \\ against accepting refugees and decided on the first speaker.\end{tabular} \\ \hline
Prepare debate & 15 min & \begin{tabular}[c]{@{}l@{}}Participants were provided with supplementary materials to \\ prepare their arguments.\end{tabular} \\ \hline
Equip sensors & 10 min & \begin{tabular}[c]{@{}l@{}}Experimenters explained wearable devices to participants and \\ assisted them in wearing devices.\end{tabular} \\ \hline
Measure baseline & 2 min & \begin{tabular}[c]{@{}l@{}}A baseline corresponding to a neutral state was measured for \\ each participant.\end{tabular} \\ \hline
Overview debate & 5 min & \begin{tabular}[c]{@{}l@{}}The moderator explained the debate rules and notified \\ participants that they are allowed to intervene.\end{tabular} \\ \hline
Debate & 10 min & \begin{tabular}[c]{@{}l@{}}Participants could speak for two consecutive minutes during \\ their turns and they were notified twice at 30 and 60 seconds \\ before the end of the debate.\end{tabular} \\ \hline
Annotate emotions & 60 min & \begin{tabular}[c]{@{}l@{}}Participants annotated emotions at intervals of every 5 \\ seconds, watching footage of themselves and their partners.\end{tabular} \\ \bottomrule
\end{tabular}
\label{table4:protocol}
\begin{tablenotes}[online]\medskip
\item[*] One session lasted approximately two hours.
\end{tablenotes}
\end{threeparttable}
\end{table}

Once they agreed to participate in the research, participants decided whether they would argue for or against admitting the Yemeni refugees in Jeju. Participants could either briefly discuss with each other to settle on their preferred positions or toss a coin to decide at random. The same procedure was followed for deciding who goes first in the debate.

Next, participants were given up to 15 minutes to prepare their arguments. Each participant was given a pen, paper, and prints of the articles that they previously received via email. After they finished preparing, experimenters equipped participants with wearable devices. Participants wore E4 wristbands on their non-dominant hand, as arm movements may impede an accurate measurement of PPG. Experimenters assured that wristbands are tightly fastened, and electrodes are in good contact with participants' skin. Experimenters also assured the EEG headsets and head-mounted cameras are well fitted on participants' head, and manually adjusted head-mounted cameras' lens to make sure the captured views are similar to participants' subjective view. Participants wore Polar H7 sensors attached to flexible bands underneath their clothes, so the electrodes are in contact with their skin and placed the sensors above their solar plexus.

\paragraph{\textit{Baseline measurement}} With all devices equipped, sensor measurements were taken from participants while they watched a short clip. This step was to establish a baseline that constitutes a neutral state for each participant. Establishing a neutral baseline is commonly used in the construction of emotion datasets to account for individual biases and reduce the effect of previous emotional states, especially when repeated measurements are taken.

A procedure for a baseline measurement varies across researchers and is often dependent on the purpose of an experiment~\cite{diers2014instructions}. In stimuli-based experiments, researchers take measurements as their subjects watch a stimulus intended to induce a neutral emotional state~\cite{soleymani2011multimodal,koelstra2011deap} or measure resting-state activities between stimuli if they are taking multiple consecutive measurements~\cite{abadi2015decaf}. Similarly, for K-EmoCon, participants watched \textit{Color Bars} clip, which was previously reported in the work of Gross et al. to induce a neutral emotion~\cite{gross1995emotion}. Experimenters also ensured that no devices were malfunctioning during the baseline measurement.

\paragraph{\textit{Debate}} A debate began at the sign of the moderator and lasted approximately 10 minutes. Participants' facial expressions, movements in their upper body, and speeches were recorded throughout a debate. Participants were allowed to speak consecutively up to two minutes during their turns, with turns alternating between two participants. However, participants were also notified that they could intervene during an opponent's turn, to allow a more natural communication. The moderator notified participants 30 and 60 seconds before the end of their turns and intervened if they exceeded two minutes. A debate stopped at the ten-minute mark with some flexibility to allow the last speaker to finish his or her argument.

\paragraph{\textit{Emotion annotation}} Participants took a 15-minute break upon finishing a debate. Participants then were each assigned to a PC and annotated their own emotions and their partner's emotions during the debate. Specifically, each participant watched two audiovisual recordings of him/herself and his/her partner from 3rd-person POV (including facial expressions, upper body movements, and speeches), to annotate emotions at intervals of every 5 seconds from the beginning to the end of a debate. We chose 5 seconds based on the report of Busso et al. that the average duration of the speaker turns in IEMOCAP was about 4.5s~\cite{busso2008iemocap}, and findings from linguistics research also support this number~\cite{kemper2001structure,yuan2006towards,gabig2013mean}.

This annotation method we employed, a \textit{retrospective affect judgment protocol}, is widely used in affective computing to collect self-reports of emotions, especially in studies where an uninterrupted engagement of subjects during an emotion induction process is essential~\cite{graesser2006detection,afzal2009natural,d2010monitoring,d2015influence}. Likewise, we opted for this method as participants' natural interaction was necessary for acquiring quality emotion data.

Note that we did not provide 1st-person POV recordings captured from head-mounted cameras to participants, and they only had 3rd-person POV recordings to annotate felt emotions. One may have a reasonable concern regarding this choice, that participants watching their faces likely caused them to occupy a perspective similar to an observer. Hence, this might have resulted in an unnatural measurement of felt emotions. Indeed, the headcam footage could have been a more naturalistic instrument, as we intuitively take an embodied perspective to recall how we felt at a specific moment in the past.

\begin{table}[!t]
\centering
\addtolength{\leftskip} {-2cm} % increase (absolute) value if needed
\addtolength{\rightskip}{-2cm}
\caption{Collected emotion annotations.}
\renewcommand{\arraystretch}{1.2}
\begin{tabular}{@{}lll@{}}
\toprule
\textbf{Emotion annotation categories} & \textbf{Description} & \textbf{Measurement scale or method} \\ \midrule
Arousal / Valence & \begin{tabular}[c]{@{}l@{}}Two affective dimensions from Russell's \\ circumplex model of affect~\cite{russell1980circumplex}\end{tabular} & \begin{tabular}[c]{@{}l@{}}1: very low - 2: low - 3: neutral \\ - 4: high - 5: very high\end{tabular} \\ \hline
\begin{tabular}[c]{@{}l@{}}Cheerful / Happy / Angry / \\ Nervous / Sad\end{tabular} & \begin{tabular}[c]{@{}l@{}}Emotion states describing a subjective \\ stress state~\cite{plarre2011continuous}\end{tabular} & \begin{tabular}[c]{@{}l@{}}1: very low - 2: low - 3: high \\ - 4: very high\end{tabular} \\ \hline
\begin{tabular}[c]{@{}l@{}}Boredom / Confusion / Delight / \\ Engaged concentration / \\ Frustration / Surprise / None\end{tabular} & \begin{tabular}[c]{@{}l@{}}Commonly used Baker Rodrigo \\ Ocumpaugh Monitoring Protocol (BROMP) \\ educationally relevant affective categories~\cite{ocumpaugh2015baker}\end{tabular} & Choose one \\ \hline
\begin{tabular}[c]{@{}l@{}}Confrustion / Contempt / Dejection / \\ Disgust / Eureka / Pride / \\ Sorrow / None\end{tabular} & \begin{tabular}[c]{@{}l@{}}Less commonly used BROMP \\ educationally relevant affective categories~\cite{ocumpaugh2015baker}\end{tabular} & Choose one \\ \bottomrule
\end{tabular}
\label{table5:annotations}
\end{table}

However, we found the extent of information captured by the headcam footage insufficient for accurate annotation of felt emotions. Experimenters manually adjusted headcam lenses, so the recordings resembled participants' subjective views, but the headcam footage was missing fine-grained information such as participants' gazes. Also, past research on memories for emotions has shown that they are prone to biases and distortion~\cite{levine2002sources,safer2002distortion,lench2010motivational}. In that regard, it seemed headcam videos, which contain limited information compared to frontal face recordings, would only result in an incorrect annotation of felt emotions, especially in retrospect. Further, we noted that it is not uncommon for people to infer emotions from their faces, as they frequently do when looking in a mirror or taking a selfie.

As a result, participants were given 3rd-person recordings of themselves for the retrospective annotation of felt emotions. In total, participants annotated emotions with 20 unique categories, as shown in Table~\ref{table5:annotations}. Experimenters assisted participants throughout the annotation procedure. Before participants began annotating, experimenters explained individual emotion categories to participants, so they correctly understood a meaning and a specific annotation procedure for each item. Experimenters also explicitly instructed participants to report felt emotions, not perceived emotions on their faces. Lastly, experimenters ensured that the start time and end time for two participants matched to obtain synchronized annotations.

% save original \intextsep
\newlength{\oldintextsep}
\setlength{\oldintextsep}{\intextsep}

\setlength\intextsep{0pt}
\begin{wraptable}{r}{4cm}
\centering
\caption{Gender and age of external raters.}
\begin{tabular}{@{}cc@{}}
\toprule
\textbf{Raters} & \textbf{Gender and age} \\ \midrule
R1 & M (27) \\
R2 & M (25) \\
R3 & F (22) \\
R4 & M (24) \\
R5 & F (28) \\ \bottomrule
\end{tabular}
\label{table6:raters}
\end{wraptable}

\paragraph{\textit{External emotion annotation}} Additionally, we recruited five external raters to annotate participants' emotions during debates (see Table~\ref{table6:raters}). We applied the same criteria we used for recruiting participants in data collection for the recruitment of the raters. The raters were provided the audiovisual debate footage of participants and annotated emotions following the same procedure our participants followed. External raters performed their tasks independently, and the experimenters communicated remotely with the raters. Once a rater finished annotating, an experimenter checked completed annotations for incorrect entries and requested a rater to review annotations if there were any missing values or misplaced entries.

% ----------------------------------------------------------------
\section{Data Records}
\subsection{Dataset summary}

The resulting K-EmoCon dataset contains multimodal data from 16 paired-debates on a social issue, which sum to 172.92 minutes of dyadic interaction. It includes physiological signals measured with three wearable devices, audiovisual recordings of debates, and continuous annotations of emotions from three distinct perspectives of the subject, the partner, and the external observers. Table~\ref{table7:summary} summarizes data collection results and dataset contents.

\setlength{\intextsep}{\oldintextsep} % restore the original \intextsep
\begin{table}[!t]
\centering
\caption{Summary of data collection results and the dataset.}
\renewcommand{\arraystretch}{1.2}
\begin{tabular}{@{}ll@{}}
\toprule
\multicolumn{2}{c}{\textbf{Data collection summary}} \\ \midrule
Number of participants & 32 (20 males and 12 females) \\ \midrule
Participants age & 19 to 36 (mean = 23.8 years, stdev. = 3.3 years) \\ \midrule
Session duration & Total 172.92 min, (mean = 10.8 min, stdev. = 1.04 min) \\ \midrule
Emotion annotations categories & \begin{tabular}[c]{@{}l@{}}\textbf{1 - 5}: Arousal, Valence\\ \textbf{1 - 4}: Cheerful, Happy, Angry, Nervous, Sad\\ \textbf{Choose one}: Common BROMP affective categories + \\ less common BROMP affective categories\end{tabular} \\ \midrule
Measured physiological signals & \begin{tabular}[c]{@{}l@{}}3-axis Acc. (32Hz), BVP (64Hz), EDA (4Hz), heart rate\\  (1Hz), IBI (n/a), body temperature (4Hz), EEG (8 band,\\  32Hz), ECG (2Hz)\end{tabular} \\ \toprule
\multicolumn{2}{c}{\textbf{Dataset contents}} \\ \midrule
Debate audios & 172.92 min (from 16 debate sessions) \\ \midrule
Debate footage & 223.35 min (from 21 participants) \\ \midrule
Physiological signals & Refer to \textit{Dataset contents} subsection \\ \midrule
\begin{tabular}[c]{@{}l@{}}Emotion annotations \\ (\# of 5-second \\ intervals annotated)\end{tabular} & \begin{tabular}[c]{@{}l@{}}\textbf{Self}: 4,159\\ \textbf{Partner}: 4,159\\ \textbf{5 external observers}: 20,803\end{tabular} \\ \bottomrule
\end{tabular}
\label{table7:summary}
\end{table}

\paragraph{\textit{Preprocessing}} For the timewise synchronization across data, we converted all timestamps from Korea Standard Time (UTC +9) to UTC +0 and clipped raw data such that only parts of data corresponding to debates and baseline measurements are included. For debate audios and the footage, subclips corresponding to debates were extracted from the raw footage. Audio tracks containing participants' speeches were copied and saved separately as WAV files. Physiological signals were clipped from the respective beginnings of data collection sessions to the respective ends of debates, as the initial 1.5 to 2 minutes immediately after a session begins corresponds to a baseline measurement for a neutral state. Parts in between baseline measurements and debates correspond to debate preparations, which may be excluded from the analysis.

% ----------------------------------------------------------------
\subsection{Dataset contents}

The K-EmoCon dataset~\cite{cheul_young_park_2020_3814370} is available upon request on \textit{Zenodo}: \url{https://doi.org/10.5281/zenodo.3814370
}. In the following, we describe directories and files in the dataset and their contents.

\paragraph{\textit{metadata.tar.gz:}} includes files with auxiliary information about the dataset. Included files are:
\begin{enumerate}
	\item \texttt{subjects.csv} -- each row contains a participant ID (\texttt{pid}) and three timestamps in UTC +0. Three timestamps respectively mark the beginning of a data collection (\texttt{initTime}), the start of a debate (\texttt{startTime}), and the end of a debate (\texttt{endTime}).
	\item \texttt{data\_availability.csv} -- shows files available for each participant. For each participant (row), if a data file (column) is available, the corresponding cell is marked \texttt{TRUE}, otherwise \texttt{FALSE}.
\end{enumerate}

\paragraph{\textit{data\_quality\_tables.tar.gz:}} includes seven CSV tables with information regarding the quality of physiological signals in the dataset. With participant IDs (\texttt{pid}) in rows and file types (\texttt{ACC}, \texttt{BVP}, \texttt{EDA}, \texttt{HR}, \texttt{IBI}, and \texttt{TEMP} for E4 data, and \texttt{Attention}, \texttt{BrainWave}, \texttt{Meditation}, and \texttt{Polar\_HR} for NeuroSky + Polar H7 data) in columns, included files are as follows:
\begin{enumerate}
	\item \texttt{e4\_durations.csv} -- contains the duration of each file in seconds, where \texttt{duration} = (last timestamp - first timestamp) / 1000.
	\item \texttt{neuro\_polar\_durations.csv} -- same as above.
	\item \texttt{e4\_zeros.csv} -- contains the number of zero values in each file. \texttt{ACC} and \texttt{BVP} were excluded as zero crossings are to be expected during their measurement.
	\item \texttt{neuro\_polar\_zeros.csv} -- same as above. Note that zero values for NeuroSky data (\texttt{Attention}, \texttt{BrainWave}, \texttt{Mediation}) indicate the inability of a device at a given moment to obtain a sufficiently reliable measurement due to various reasons.
	\item \texttt{e4\_outliers.csv} -- contains the number of outliers in each file. Chauvenet's criterion was used for outlier detection (refer to \textit{Code Availability} section for its implementation in Python).
	\item \texttt{e4\_completeness.csv} -- contains the completeness of each file as a ratio in the range of [0.0, 1.0]. 1.0 indicates a file without any missing value or an outlier. The completeness ratio was calculated as \texttt{completeness} = (total number of values - (number of outliers + number of zeros)) / total number of values.
	\item \texttt{neuro\_polar\_completeness.csv} -- same as above, with completeness calculated as \texttt{completeness} = (total number of values - number of zeros) / total number of values.
\end{enumerate}

\paragraph{\textit{debate\_audios.tar.gz:}} contains 16 audio recordings of debates in the WAV file format. The name of each file follows the convention of \texttt{p<X>.p<Y>.wav}, where \texttt{<X>} and \texttt{<Y>} stand for IDs of two participants appearing in the audio. The start and the end of each recording correspond to \texttt{startTime} and \texttt{endTime} values in the subjects.csv file, respectively.

\paragraph{\textit{debate\_recordings.tar.gz:}} contains video recordings of 21 participants during debates in the MP4 file format. The name of a file \texttt{p<X>\_<T>.mp4} indicates that the file is the recording of participant \texttt{<X>} that is \texttt{<T>} seconds long.

\paragraph{\textit{neurosky\_polar\_data.tar.gz:}} includes subdirectories for each participant, from P1 to P32, which may contain up to four files as the following:
\begin{enumerate}
	\item \texttt{Attention.csv} --  contains \textit{eSense Attention} ranging from 1 to 100, representing how attentive a user was at a given moment. Attention values can be interpreted as the following: 1 to 20 -- ``strongly lowered'', 20 to 40 -- ``reduced'', 40 to 60 -- ``neutral'', 60 to 80 -- ``slightly elevated'', and 80 to 100 -- ``elevated''. 0 indicates that the device was unable to calculate a sufficiently reliable value, possibly due to a signal contamination with noises.
	\item \texttt{BrainWave.csv} -- records the relative power of brainwave in the following 8 bands of EEG: delta (0.5 - 2.75Hz), theta (3.5 - 6.75Hz), low-alpha (7.5 - 9.25Hz), high-alpha (10 - 11.75Hz), low-beta (13 - 16.75Hz), high-beta (18 - 29.75Hz), low-gamma (31 - 39.75Hz), and middle-gamma (41 - 49.75Hz). The values are without a unit and are only meant for inferring the fluctuation in the power of a certain band or comparing the relative strengths of bands with each other.
	\item \texttt{Meditation.csv} -- contains \textit{eSense Meditation} ranging from 0 to 100, measuring the relaxedness of a user. For their interpretation, use the same ranges and the meanings as those for the attention values.
	\item \texttt{Polar\_HR.csv} - contains heart rates measured with ECG sensors during debates.
\end{enumerate}

\paragraph{\textit{e4\_data.tar.gz:}} contains subdirectories for each participant (except P2, P3, P6, and P7), which may contain up to six files as the following:
\begin{enumerate}
	\item \texttt{E4\_ACC.csv} -- measurements from a 3-axis accelerometer sampled at 32Hz in the range [-2g, 2g] under columns x, y, and z. Multiply raw numbers by 1/64 to convert them into units of g (i.e., a raw value of 64 is equivalent to 1g).
	\item \texttt{E4\_BVP.csv} -- PPG measurements sampled at 64Hz. 
	\item \texttt{E4\_EDA.csv} -- EDA sensor readings in units of $\mu$S, sampled at 4Hz.
	\item \texttt{E4\_HR.csv} -- the average heart rates calculated in 10-second windows. The values are derived from the BVP measurements, and the values are entered at the frequency of 1Hz. The first 10 seconds of data after the beginning of a recording is not included as the derivation algorithm requires the initial 10 seconds of data to produce the first value.
	\item \texttt{E4\_IBI.csv} -- IBI measurements in milliseconds computed from the BVP. From a second row onwards, one row is separated from the previous row with an amount equal to a distance between two peaks (i.e., $t_{i+1} - t_{i} = IBI_{i}$). Note that HR in terms of BPM can be derived from IBI by taking $60/(IBI * 1000)$.
	\item \texttt{E4\_TEMP.csv} -- a body temperature measured in the Celsius scale at the frequency of 4Hz.
\end{enumerate}

Note that E4 data entries for P29, P30, P31, and P32 are entered with each row designated with either one of two unique \texttt{device\_serial} values. It is necessary that users of this dataset only use rows corresponding to a single \texttt{device\_serial}. We further recommend using rows with the following \texttt{device\_serial} values:
\begin{enumerate}
	\item[$\bullet$] P29, P31 -- \texttt{A013E1} for all files, except \texttt{A01525} for IBI.
	\item[$\bullet$] P30, P32 -- \texttt{A01A3A} for all files.
\end{enumerate}

\paragraph{\textit{emotion\_annotations.tar.gz:}} includes four subdirectories as listed below, which each contain annotations for participant emotions during debates at intervals of every 5 seconds, acquired from three distinct perspectives:
\begin{enumerate}
	\item \texttt{self\_annotations} -- annotations of participant emotions by participants themselves.
	\item \texttt{partner\_annotations} -- annotations of participant emotions by respective debate partners.
	\item \texttt{external\_annotations} -- annotations of participant emotions by five external raters. Files follow the naming convention of \texttt{P<X>.R<Z>.csv}, where \texttt{<X>} is a participant ID, and \texttt{<Z>} is a rater number.
    \item \texttt{aggregated\_external\_annotations} -- contains external rater annotations aggregated across five raters via majority voting. Refer to \textit{Code Availability} section for the Python code implementing the majority vote aggregation.
\end{enumerate}

The first row in a valid file has annotations for the first five seconds, and rows coming afterward contain annotations for the next consecutive five-second intervals,  non-overlapping with each other. Also, each row in a valid file contains 10 non-empty values (eight numeric values, including \texttt{seconds} column, and two x's). Note that annotation files for a participant may not have an equal number of rows (e.g., there may be more self-annotations than partner/external annotations for some participants). In that case, longer files should be truncated from the start such that they have the same number of rows as shorter files since the extra annotations at the beginning are possibly from participants mistakenly annotating emotions during baseline measurements.

% ----------------------------------------------------------------
\section{Technical Validation}
\subsection{Emotion annotations}

\begin{figure}[!t]
\centering
\includegraphics[width=\textwidth]{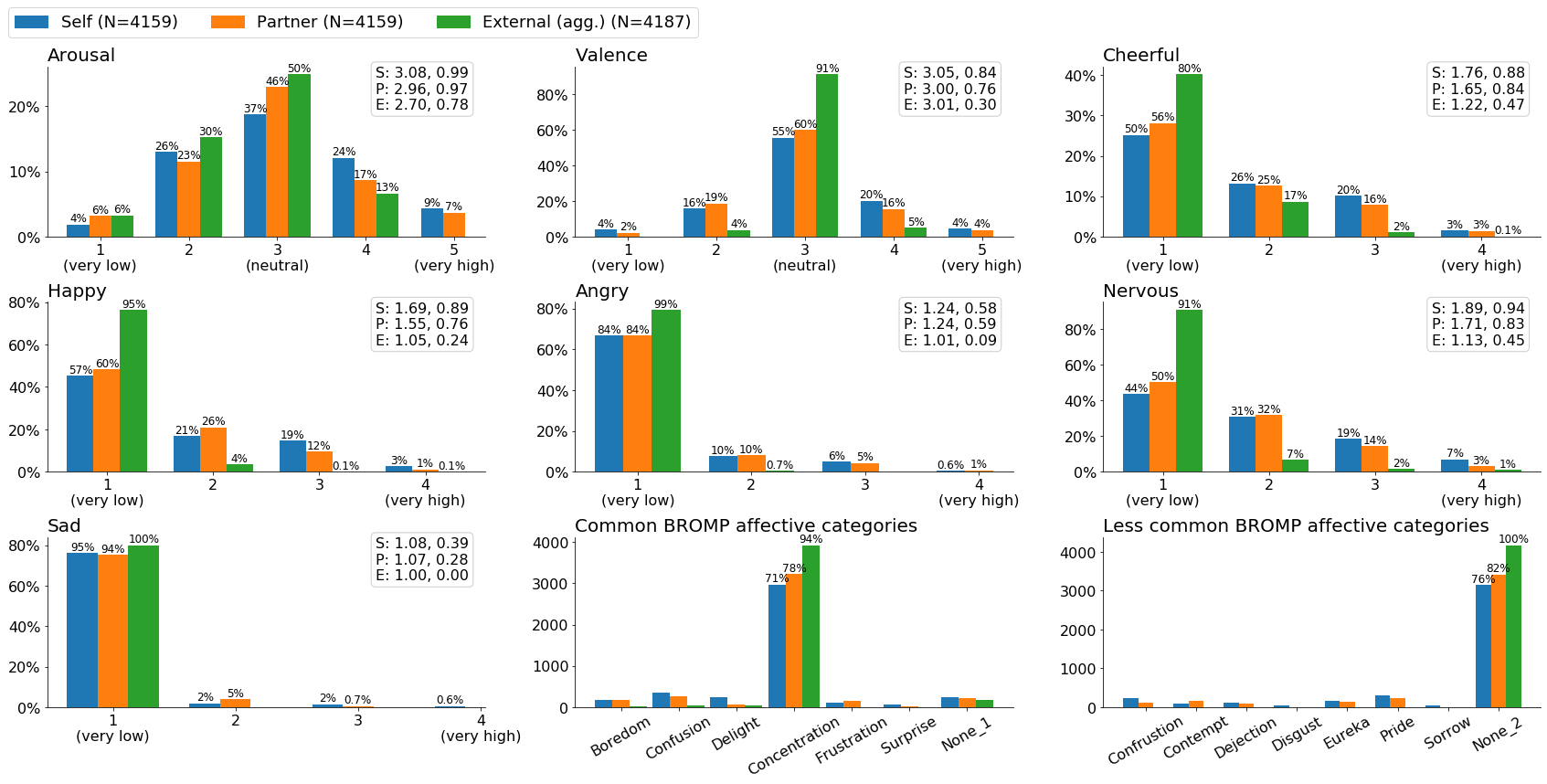}
\caption{Distributions and frequencies of emotion annotations from three perspectives of self (S), partner (P), and external raters (E), with external annotations aggregated by majority voting. Annotations were summed across 32 subjects for each emotion and affective categories. Means and standard deviations measured respectively from three perspectives are shown on the upper right corner of figures if available.}
\label{fig3:dist_and_freq}
\end{figure}

\paragraph{\textit{Distribution and frequency of emotions}} The distributions and the frequencies of emotion annotations are as shown in Figure~\ref{fig3:dist_and_freq}. Overall, annotations for emotions measured on Likert scales (arousal, valence, cheerful, happy, angry, nervous, and sad) are biased towards a neutral with only a minuscule fraction of annotations for non-neutral states. Categorical emotion annotations (common and less common BROMP affective categories) are similarly biased, with a predominant portion of annotations falling under only two categories of concentration and none. This imbalance in annotations is as expected as emotion data is commonly imbalanced by its nature in the wild (i.e., people are more often neutral than angry or sad)~\cite{calix2010emotion,wang2012harnessing,xu2015word}.

\begin{figure}[!t]
\centering
\includegraphics[width=\textwidth]{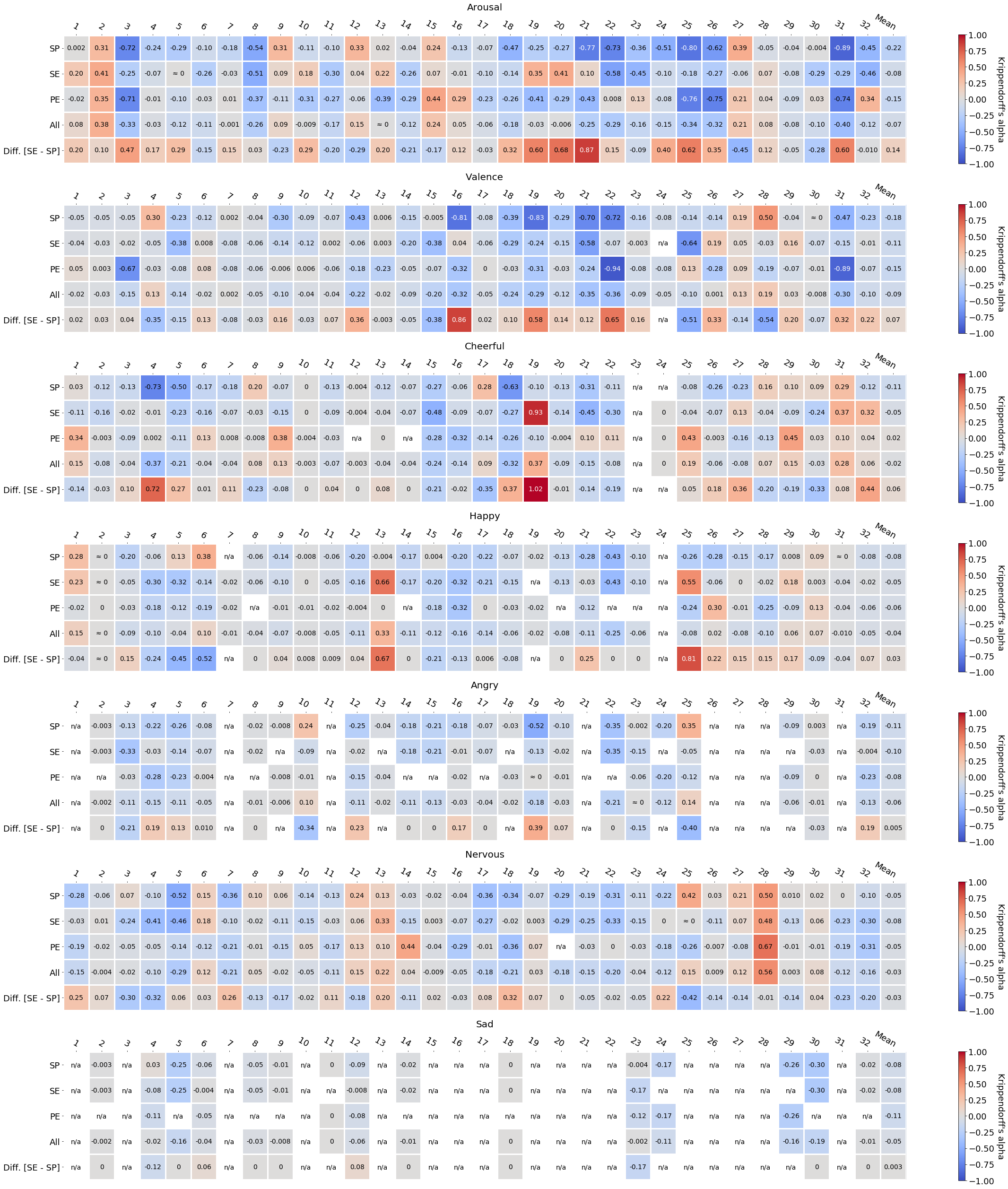}
\caption{Heatmaps of inter-rater reliabilities measured with Krippendorff's alpha. External annotations were aggregated by majority voting. The first 4 rows of each heatmap show alpha coefficients across four different combinations of annotation perspectives: (1) \textit{SP} = self vs. partner, (2) \textit{SE} = self vs. external, (3) \textit{PE} = partner vs. external, and (4) \textit{All} = self vs. partner vs. external, while the last row (\textit{Diff [SE - SP]}) shows the difference between self vs. external agreement and self vs. partner agreement. The columns show those for each participant.}
\label{fig4:irr_heatmaps}
\end{figure}

\paragraph{\textit{Inter-rater reliability}} As individual-level information is missing in aggregated data, we used Krippendorff's alpha~\cite{krippendorff2011computing}, which is a generalized statistic of agreement applicable to any number of raters, to measure the inter-rater reliability (IRR) of emotion annotations from different perspectives for each participant. Figure~\ref{fig4:irr_heatmaps} shows heatmaps of alpha coefficients computed for seven emotions measured on ordinal scales (arousal, valence, cheerful, happy, angry, nervous, and sad).

All annotation values were interpreted as rank-ordered (ordinal scaled) for the IRR computation. Likert scales we used are not intervals or ratios with meaningful distances in-between. While participants and raters were provided numeric scales labeled with semantic meanings (see Table~\ref{table5:annotations}), the individual interpretations of scales were likely disparate.

Given that, before the computation, annotation values were scaled relative to a neutral, by estimating modes of columns as neutrals and deducting them from respective column values (i.e., if the mode of a \textit{cheerful} column for a particular participant was one, then one was subtracted from all values in that \textit{cheerful} column). This \textit{mode-subtraction} step was necessary to prevent the underestimation of IRRs.

Annotations in our dataset for scaled emotions are highly biased as shown in Fig.~\ref{fig3:dist_and_freq}). However, while arousal and valence are explicitly centered at zero (which corresponds to 3 = neutral), five emotions measured in the scale of 1 = \textit{very low} to 4 = \textit{very high} (cheerful, happy, angry, nervous, and sad) are systematically biased without a zero neutral. All of their values indicate that some emotion is present, and this absence of zero results in a widely varying interpretation of scale values by our participants and raters.

Consider the following scenario further elaborating this issue: a subject rates that she was cheerful as much as 1 for the first third of a debate, then 2 for the rest, but her debate partner rates that she was cheerful as much as 3 for the first third then 4 for the rest. In this example, self and partner annotations both imply that the subject was less cheerful for the first third of the debate. However, an IRR of two sets of annotations is close to zero without subtracting modes. Indeed, it is possible that the partner perceived the subject as more cheerful overall, compared to the subject herself. In that case, a low IRR correctly measures the difference between emotion perceptions of the subject and partner. Nevertheless, this assumption cannot be confirmed, as there is no neutral baseline.

Therefore, we applied the proposed mode-subtraction to emotion annotations such that alpha coefficients measure raters' agreement on relative changes in emotions rather than their absolute agreement with each other. This adjustment mitigates spuriously low alpha coefficient values obtained from raw annotations (refer to \textit{Code Availability} section for the code implementing the mode-subtraction and plotting of heatmaps).

These fixed alpha coefficients are low in general. In particular, a noticeable pattern emerges when comparing alpha coefficients of self-partner (\textit{SP}) annotations and self-external (\textit{SE}) annotations. As shown in the last rows of heatmaps \textit{(Diff. [SE - SP])} in Fig.~\ref{fig4:irr_heatmaps}, the differences between the IRRs of SE annotations and SP annotations tend to be above zero (for 20 out of 32 participants for arousal: mean = 0.143, stdev. = 0.322). This pattern possibly indicates that there exists a meaningful difference in the perception of emotions from different perspectives, while further study is required to validate its significance.

% ----------------------------------------------------------------
\subsection{Physiological signals}

\paragraph{\textit{Data quality}} The quality of physiological signal measurements in the dataset has been thoroughly examined. The examination results are included as a part of the dataset in the \texttt{data\_quality\_tables.tar.gz} archive file.

\paragraph{\textit{Missing data}} E4 data of 4 participants (P2, P3, P6, and P7) were excluded due to a device malfunction during data collection. While physiological signals in the dataset are mostly error-free with most of the files complete above 95\%, a portion of data is missing due to issues inherent to devices or a human error:
\begin{enumerate}
	\item[$\bullet$] IBI -- data from P26 is missing as the internal algorithm of E4 that derives IBI from BVP automatically discards an obtained value if its reliability is below a certain threshold.
	\item[$\bullet$] EDA -- data from P17 and P20 is missing, possibly due to poor contact between the device and a participant's skin.
	\item[$\bullet$] NeuroSky (Attention, Meditation) -- measurements from P1 and P20 are missing due to a poorly equipped device. A portion of data is missing for P19 (~32\%), P22 (~59\%) and P23 (~36\%) for the same reason. No BrainWave data was lost.
	\item[$\bullet$] Polar HR -- data from seven participants (P3, P12, P18, P20, P21, P29, and P30) are missing due to a device error during data collection. Parts of data are missing from P4 (~38\%) and P22 (~38\%) due to poor contact.
\end{enumerate}

% ----------------------------------------------------------------
\section{Usage Notes}
\subsection{Potential applications}

In addition to the intended usage of the dataset discussed above, there are uncertainties as to how physiological markers of an individual's capacity for flexible physiological reactivity relate to experiences of positive and negative emotions. Our dataset could potentially be useful to examine the role of physiological signal based markers in assessing an individual's use of emotion regulation strategies, such as cognitive appraisal.

Additionally, various data mining and machine learning techniques could be applied to set up the models for an individual's emotional profiles based on sensor-based physiological and behavioral recordings. This could further be transferred to other use-cases, such as helping children with autism in their social communication~\cite{picard2009future,washington2017superpowerglass}, helping people who are blind to read facial expressions and get the emotion information of their peers~\cite{buimer2018conveying}, helping robots interact more intelligently with people~\cite{breazeal2003emotion,kwon2007emotion}, and monitoring signs of frustration and emotional saturation that affect attention while driving, to enhance driver safety~\cite{nass2005improving,eyben2010emotion}.

% ----------------------------------------------------------------
\subsection{Limitations}

\paragraph{\textit{Data collection apparatus}} Contact-base EEG sensors are known to be susceptible to noises, for example, frowning or eyes-movement might have caused peaks in the data. Other devices may also have been subject to similar systematic errors.

\paragraph{\textit{Data collection context}} The context of the turn-taking debate may have caused participants to regulate or even suppress their emotional expressions, as an unrestrained display of emotions is often regarded undesirable during a debate. This may have contributed to a deflated level of agreement between self-reports and partner/external perceptions of emotions, which may not be a case for more natural interactions in the wild.

\paragraph{\textit{Demographics}} The participant demographics likely have introduced bias in the data. All of our participants and raters are young (their ages were between 19 to 36) and highly-educated, and the majority of them are individuals of Asian ethnicity. Therefore, our data may not generalize well to individuals of different ethnic groups or of younger or older age groups.

\paragraph{\textit{Unaccounted variables}} Many variables unaccounted during data collection, such as the level of rapport between debating pairs, a participant's competence in spoken English, and a participant's familiarity with the debate topic, may also have contributed to a variance in the level of mismatch between the perceptions of emotions across different perspectives.

% ----------------------------------------------------------------
\section{Code Availability}

Python codes implementing outlier detection using Chauvenet's criterion, majority voting, mode-subtraction, and other utility functions, including the generation of heatmap plots, are available on \url{https://github.com/Kaist-ICLab/K-EmoCon_SupplementaryCodes}. The \textit{Krippendorff} Python package (\url{https://github.com/pln-fing-udelar/fast-krippendorff}) was used for the computation of Krippendorff's alpha. The Python of version 3.6.9 was used throughout.

Codes for preprocessing the raw log-level data in SQL databases to CSV files were implemented in Python with the \textit{SQLAlchemy} package. However, these codes and the raw log-level data are not made available as they include privacy-sensitive information outside the agreed boundary for public sharing of the dataset, which was collected only for logistic reasons. Nevertheless, we welcome users of the dataset to contact the corresponding authors if they need any further assistance or information regarding the raw data, and it's preprocessing.

% ----------------------------------------------------------------
\bibliographystyle{unsrt}
% \bibliography{references}  %%% Remove comment to use the external .bib file (using bibtex).


\end{document}